\documentclass[%
 aip,
 amsmath,amssymb,
 reprint,%
]{revtex4-1}

\usepackage{graphicx}% Include figure files
\usepackage{dcolumn}% Align table columns on decimal point
\usepackage{bm}% bold math
\usepackage[utf8]{inputenc}
\usepackage[T1]{fontenc}
\usepackage{mathptmx}
\usepackage{etoolbox}

\usepackage[colorlinks=true, linkcolor=blue, citecolor=blue, urlcolor=black]{hyperref}

\hypersetup{
    linkcolor=blue,  % Color of links (title, sections, etc.)
    citecolor=blue,  % Color of citations
    urlcolor=blue   % Color of URL links
}

\makeatletter
\def\@email#1#2{%
 \endgroup
 \patchcmd{\titleblock@produce}
  {\frontmatter@RRAPformat}
  {\frontmatter@RRAPformat{\produce@RRAP{*#1\href{mailto:#2}{#2}}}\frontmatter@RRAPformat}
  {}{}
}%
\makeatother
\begin{document}

\preprint{AIP/123-QED}

% Make the title itself a hyperlink in blue
%\title{\href{http://www.example.com}{Quantum Well Effects on Tunnel Electroresistance and Resistance-Area Product in Ferroelectric Tunnel Junctions}}
\title{Asymmetric Resonant Ferroelectric Tunnel Junctions for Simultaneous High Tunnel Electroresistance and Low Resistance-Area Product}
% \title[]{Integrated Quantum Transport Modelling of Ferroelectric Tunnel Junction}

% Force line breaks with \\
\author{Balram Khattar}
% \affiliation{ Department of Electrical Engineering, Indian Institute of Technology Ropar, Rupnagar 140001, India }  %\\This line break forced with \textbackslash\textbackslash}%

\author{Adarsh Tripathi}%
\affiliation{ Department of Electrical Engineering, Indian Institute of Technology Ropar, Rupnagar 140001, India}

\author{Manmohan Brahma}
% \email{abhi@iitmandi.ac.in}
\affiliation{% 
Academy of Scientific and Innovative Research (AcSIR), 201002, Ghaziabad, India %\\This line break forced% with \\
}%

\author{Abhishek Sharma}
\email{abhi@iitmandi.ac.in}
\affiliation{% 
School of Computing and Electrical Engineering, Indian Institute of Technology Mandi, Mandi 175005, India %\\This line break forced% with \\
}%

\date{\today}% It is always \today, today,
             %  but any date may be explicitly specified
%%%%%%%%%%%%%%%%%%%%%%%%%%%%%%%%%%%%%%%%%%%%%%%%%%%%%%%%%%%%%%%%%%%%%%%%%%%%%%%%%%%%%%%%
\begin{abstract}
    \noindent
    \begin{center}
        \begin{minipage}{0.85\textwidth} % Adjust the width as necessary
    Ferroelectric tunnel junctions offer potential for non-volatile memory with low power, fast switching, and scalability, but their performance is limited by a high resistance-area product and a low tunnel electroresistance ratio. To address these challenges, we propose a doped $\mathrm{HfO_2}$-based, silicon-compatible asymmetric resonant ferroelectric tunnel junction design with a quantum well embedded between two ferroelectric layers, replacing the conventional metal-ferroelectric-metal structure. Using a self-consistent coupling of the non-equilibrium Green’s function method with a Preisach-based model, we demonstrate that the quantum well enhances resonant tunneling effects, leading to a simultaneous reduction in the resistance-area product and a boost in the tunnel electroresistance ratio. The low-resistance state becomes more robust, while the high-resistance state is suppressed, improving readout speed and reducing power usage. We observed that incorporating a 2\,nm quantum well significantly enhances the tunnel electroresistance ratio, achieving a peaking value of approximately $6.15 \times 10^{4}$\%, while simultaneously minimizing the resistance-area product to $4.71 \times 10^{1}$~$\Omega$-$\mathrm{cm}^2$ at 0.175\,V. Additionally, the device exhibits negative differential resistance, further enhancing its functionality. Our results confirm that this design enables scalable, energy-efficient, and high-performance non-volatile memory, making it a strong candidate for future memory technologies.
        \end{minipage}
    \end{center}
\end{abstract}
\maketitle

Ferroelectric Tunnel Junctions (FTJs) are emerging as key components in next-generation electronic devices due to their ability to modulate resistance states using an ultrathin ferroelectric (FE) barrier.\cite{park2024ferroelectric, Kohlstedt2005} These devices consist of two different conductive electrodes that are sandwiched between a thin layer of FE material, acting as the barrier for tunneling. The electrodes screening length leads to polarization-dependent band bending, which modifies the effective barrier height and, in turn, influences the tunneling electroresistance (TER).\cite{luo2025review} The change in resistance, crucial for FTJ operation, is directly governed by the reversal of polarization in the FE layer, leading to two distinct resistance states: a low-resistance state (LRS) and a high-resistance state (HRS). The TER ratio, which quantifies this effect, is defined as: $\mathrm{TER~ratio~(in~\%) = [(LRS- HRS)/LRS ]\times 100}$.\cite{Kobayashi2019} These unique properties make FTJs highly suitable for applications in non-volatile memory, logic circuits, and neuromorphic computing.\cite{covi2022ferroelectric, hwang2024smallreview} Their ability to store data in distinct resistance states provides an energy-efficient alternative to traditional memory technologies, paving the way for advancements in high-performance, low-power electronic devices.\cite{jao2021design}\\
The integration of doped $\mathrm{HfO_2}$ into existing semiconductor fabrication processes enables seamless scalability for practical applications.\cite{goh2020ultra, ambriz2017tunneling, max2020hafnia} However, implementing doped $\mathrm{HfO_2}$-based FTJs in crossbar arrays based memory design presents challenges, particularly in readout performance.\cite{jao2021design, athle2024ferroelectric} Addressing these challenges requires enhancing the TER ratio while maintaining a low resistance-area (RA) product, enabling FTJs with enhanced energy efficiency, device miniaturization, high packing density, and long-term reliability. A high TER ratio is crucial for clear resistance state differentiation, which is essential for reliable data storage, accurate readout, and robust logic operations. Simultaneously, a low RA product particularly important for scaling down device dimensions while maintaining high packing density in FTJ-based memory arrays. As devices shrink, a high TER ratio ensures stable operation, while a low RA prevents excessive resistance that could hinder current flow. Balancing these factors is key to developing ultra-low-power, high-speed, and high-density non-volatile memory architectures.\\
Various microscopic processes influence the tunneling barrier in FTJs, including those occurring at the interfaces, within the electrodes, and inside the FE layer. Asymmetry in these structures—achieved through careful selection of electrodes \cite{Kobayashi2019, Yoon2019}, composite barrier layers \cite{ wang2016overcoming}, and interface engineering \cite{lu2014ferroelectric}—plays a crucial role in enhancing the TER ratio. To optimize the TER ratio and RA product, several studies have explored different FTJ configurations. Zhuravlev \textit{et al.} (2005, 2009) conducted theoretical analyses metal-FE-metal (MFM) stack 
($\mathrm{SrRuO_3/PbZr_{0.52}Ti_{0.48}O_3/Pt}$) and metal-FE-IL-metal (MFIM) structures incorporating $\mathrm{BaTiO_3}$ as the FE layer and $\mathrm{SrTiO_3}$ as the nonpolar dielectric. Their theoretical models predicted significant enhancements in the TER ratio while offering a considerable RA product at large TER ratio values.\cite{zhuravlev2005giant,Zhuravlev2009} Expanding on this, Chang \textit{et al.} (2017) conducted theoretical calculations on MFM and MFIM stacks using inorganic ($\mathrm{BaTiO_3}$) and organic (PVDF, polyvinylidene fluoride) FE materials, demonstrating a strong correlation with experimental results.\cite{Chang2017pra} By employing the Landau-Khalatnikov approach integrated with the nonequilibrium Green’s function (NEGF) formalism, they observed a relatively low TER ratio, highlighting the need for further optimization. Similarly, Zheng \textit{et al.} (2022) investigated resonant band engineering in FTJs, evaluating both RA and TER ratio parameters. Using perovskite-based $\mathrm{BaTiO_3}$ FEs with a $\mathrm{BaSnO_3}$ dielectric layer, they demonstrated polarization-dependent RT, achieving a substantial TER with a low RA product.\cite{zheng2022band} While these studies have shown promising results, a common limitation remains—the inherent incompatibility of perovskite-based FEs with CMOS technology, posing challenges for practical integration. 
%Su \textit{et al.} (2021) explored resonant tunneling (RT) in FTJs through first-principles modeling based on density functional theory.\cite{su2021prb} Their study focused on a $\mathrm{SrRuO_3/BaTiO_3/SrRuO_3}$ FTJ with a $\mathrm{BaSnO_3}$ monolayer embedded within the $\mathrm{BaTiO_3}$ barrier, leveraging perovskite FEs to enable RT.
% In a different approach, Su \textit{et al.}~(2022) investigated RT in FTJs using first-principles calculations based on density functional theory and a quantum mechanical tunneling model. They analyzed three FTJ structures: (1) an asymmetry-induced TER-based FTJ, (2) a resonant band-induced TER-based FTJ, and (3) an advanced RT-based FTJ integrating both mechanisms to enhance TER while reducing the resistance-area product. The third FTJ, FTJ-3, utilized a $\mathrm{SrRuO_3/BaTiO_3/BaSnO_3/SrTiO_3/SrRuO_3}$ structure, where the inclusion of $\mathrm{BaSnO_3}$ within the $\mathrm{BaTiO_3}$ and $\mathrm{SrTiO_3}$ barrier optimized RT effects. The embedded $\mathrm{BaSnO_3}$ modulated band alignment, enabling a transition between direct and resonant tunneling, enhancing the TER effect while reducing the resistance-area product.\cite{su2022npj}
More recently, Chang \textit{et al.} (2023) explored $\mathrm{HfO_2}$-doped FEs for resonant FTJs design by integrating the Preisach-based model with the transfer matrix method. Although the proposed resonant FTJs are CMOS-compatible, they exhibit a relatively low TER ratio and a high RA product.\cite{chang2023edl}
\begin{figure}[htb!]
    \centering
    \includegraphics[scale=0.58]{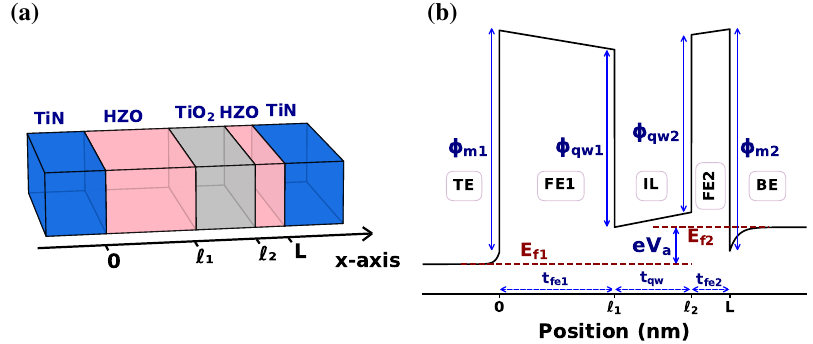}
    \caption{(a) Schematic of $\mathrm{TiN/HZO/TiO_2/HZO/TiN}$ stack FTJ and (b) the energy band diagram for multilayer FTJ with one QW.}
    \label{1stack}
\end{figure}

Hence, in this work, we develop a Preisach-based model integrated with the NEGF formalism to present a design guideline for the simultaneous enhancement of the TER ratio and reduction of the RA product through resonant band engineering, achieved via appropriate material selection and precise tuning of quantum well (QW) thickness. Our work utilizes $\mathrm{HfO_2}$-doped FEs to ensure CMOS compatibility while optimizing both TER ratio and RA product—key factors for seamless integration into practical electronic devices.
By incorporating a low-barrier insulating layer (IL) within the FE barriers, we can effectively tailor the electronic transport characteristics of FTJs. The addition of an IL, such as $\mathrm{TiO_2}$, which has been shown to improve the FE characteristics of HZO films, plays a crucial role in enhancing the TER ratio in $\mathrm{HfO_2}$-based FTJs.\cite{ hwang2020effect, joh2021stress} This IL introduces resonant states within the energy gap of the barrier, which can be modulated through FE polarization. By carefully selecting the QW thickness, the resonant states can be tuned to optimize the transition between direct and resonant tunneling transport modes, further improving the TER ratio and RA product. This precise band engineering significantly enhances the resistance ratio of FTJs, providing a practical approach to meeting industry performance standards. To explore this mechanism, we investigate RT effects in FTJs with a QW structure, employing a Preisach-based model integrated with the NEGF formalism. By leveraging this integrated modeling approach, we accurately assess the impact of QW thickness on the TER ratio and RA characteristics of FTJs, offering critical insights into optimizing barrier design and material properties for improved device efficiency.

The schematic layout of the $\mathrm{TiN/HZO/TiO_2/HZO/TiN}$ multilayer stack FTJ is depicted in \hyperref[1stack]{Fig.~\ref{1stack}(a)}. The IL $\mathrm{TiO_2}$, with a thickness of $\mathrm{t_{qw}}$, is positioned between two FE HZO ($\mathrm{Hf_{0.5}Zr_{0.5} O_2}$) layers with thicknesses $\mathrm{t_{fe1}}$ and $\mathrm{t_{fe2}}$. We have implemented the asymmetrical barrier in FTJ, utilizing the same material, TiN, for the top electrode (TE) and bottom electrode (BE). The FE thickness $\mathrm{t_{fe}}$ is the total sum of FE thicknesses $\mathrm{t_{fe1}}$ and $\mathrm{t_{fe2}}$ in the MFIFM FTJ. The corresponding energy band diagram is shown in \hyperref[1stack]{Fig.~\ref{1stack}(b)}. The IL is performing as a QW that enables RT in the MFIFM FTJ structure. We consider a Thomas-Fermi model of screening for determining the distribution of screening charge and potential profile at the metal-FE interface. The screening potential at the metal-FE interface is represented by the following Eq.~\cite{kittel2018introduction,sandu2022insights,pantel2010electroresistance}
\begin{equation}
\mathrm{V_{m1(m2)}(x)}=
\begin{cases} 
    \mathrm{-\dfrac{\sigma_s \lambda_{m1}}{\varepsilon_{m1} \varepsilon_o} e^{\left(x / \lambda_{m1}\right)}}, & \text{if~  }\mathrm{ x \leq 0}\\[10pt]
   \hspace{0.2cm} 
   \mathrm{\dfrac{\sigma_s \lambda_{m2}}{\varepsilon_{m2} \varepsilon_o} e^{\left[-\left(x-L\right)/ \lambda_{m2}\right]}}, & \text{if~  }  \mathrm{x \geq L}
    \end{cases}
    \label{thomas_fermi}
\end{equation}
where $\mathrm{\sigma_s}$ represents the surface charge density at the metal-FE interfaces, $\mathrm{\lambda_{m1}}$ ($\mathrm{\varepsilon_{m1}}$) and $\mathrm{\lambda_{m2}}$ ($\mathrm{\varepsilon_{m2}}$) denote the effective lengths of screening (relative dielectric constants) for the TE and BE electrodes, respectively, and $\mathrm{\varepsilon_o}$ signifies the permittivity of free space. The boundary conditions allow us to determine the surface charge density, which can be expressed as~\cite{Chang2017pra}
\begin{equation}
    \mathrm{\sigma_s=\varepsilon_o \varepsilon_\mathrm{fe,i} E_{\mathrm{fe,i}} + P_{fe,i}= \mathrm{\varepsilon_o \varepsilon_\mathrm{qw} E_{qw}}}
    \label{gauss_law_il_fe}
\end{equation} 
\begin{figure}[htb!]
    \centering
    \includegraphics[scale=0.23]{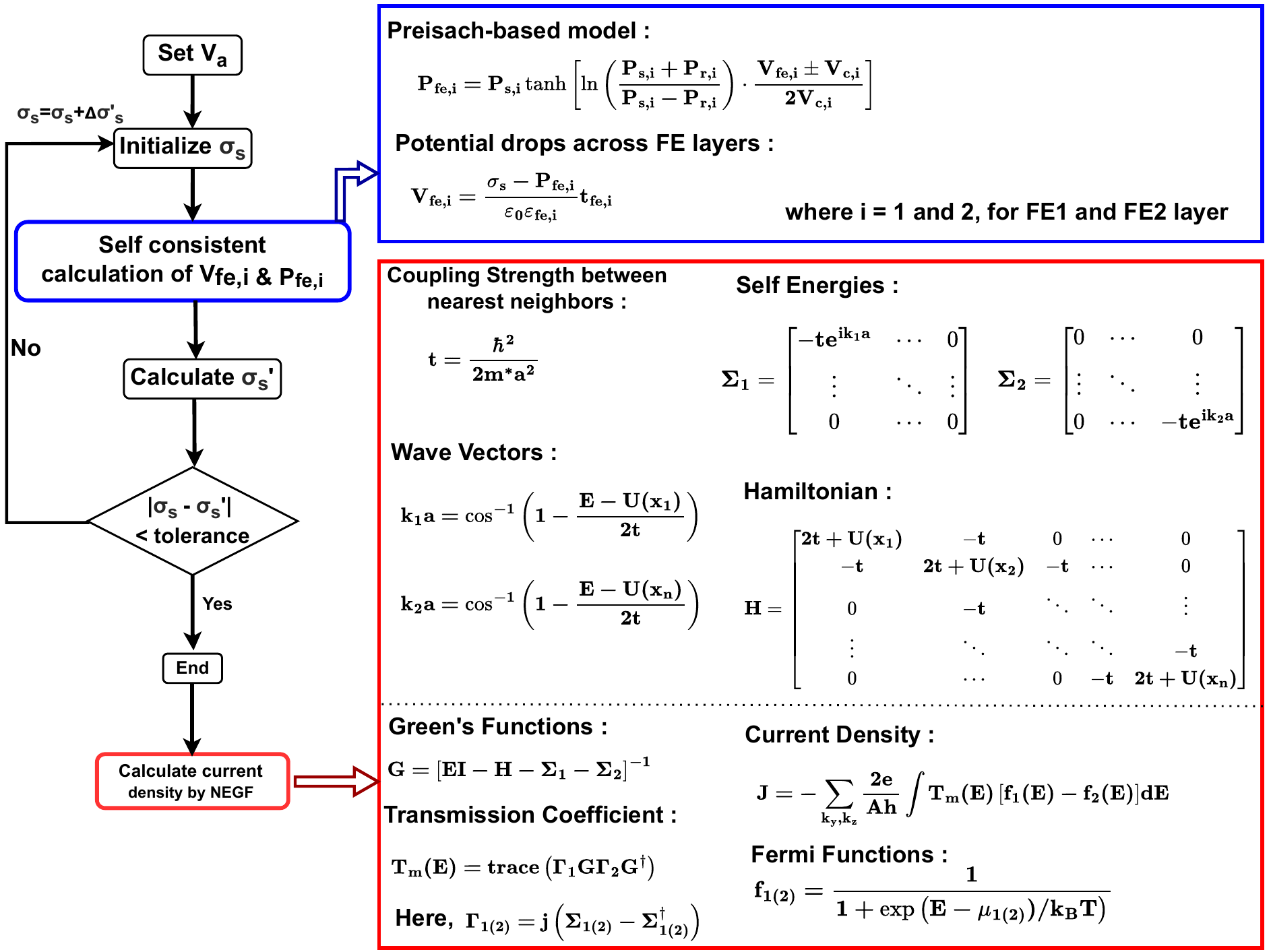}
    \caption{The computation of the self-consistent current for multilayer FTJ by using the integration of a Preisach-based model with the NEGF formalism.}
    \label{2simframe}
\end{figure}
Here, $\mathrm{i}$ designates 1 and 2 for the FE1 and FE2 layers, respectively, $\mathrm{P_{fe,i}}$, $\mathrm{E_{fe,i}}$ and $\mathrm{\varepsilon_\mathrm{fe,i}}$ are polarization, electric field, and the relative dielectric constant of $\mathrm{FE_i}$ layers. The voltage drop across this structure is determined as
\begin{equation}
    \mathrm{V_{m1}(0) + V_{m2}(L) + V_{fe1} + V_{fe2} + V_{qw} = V_a + V_{bi}}
    \label{overall_drop}
\end{equation}
where, $ \mathrm{ V_{m1}(0)=-\sigma_s \lambda_{m1}/\varepsilon_{m1}\varepsilon_o}$ and $ \mathrm {V_{m2}(L)=\sigma_s \lambda_{m2}/ \varepsilon_{m2} \varepsilon_o}$ are the potential drops at locations 0 and L, respectively. $\mathrm{V_{fe1}}$, $\mathrm{V_{fe2}}$ and $\mathrm{V_{qw}}$ are the voltage drops across the FE1, FE2 layer, and  QW, respectively. The applied voltage is represented by $\mathrm{V_{a}}$, whereas $\mathrm{V_{\mathrm{bi}}}$ indicates the built-in potential, which can be expressed as
\begin{equation}
    \mathrm{V_{bi}= \phi_{m2} - \phi_{m1} - \phi_{qw2} + \phi_{qw1}}
    \label{built_in_drop}
\end{equation}
where, $\mathrm{\phi_{m1}}$, $\mathrm{\phi_{qw1}}$, $\mathrm{\phi_{qw2}}$, and $\mathrm{\phi_{m2}}$ represent the barrier heights at the interface positions 0, $\ell_1$, $\ell_2$, and L, respectively [as illustrated in \hyperref[1stack]{Fig.~\ref{1stack}(b)}]. The total FE potential can be expressed as $\mathrm{V_{fe} = V_{fe1} + V_{fe2}}$. From Eq.~\hyperref[gauss_law_il_fe]{(\ref{gauss_law_il_fe})}, the potential drop $\mathrm{V_{qw}}$ across the QW can be represented as $\mathrm{V_{qw}=\sigma_s t_{qw}/\varepsilon_o \varepsilon_{qw}}$, with $\mathrm{\varepsilon_{qw}}$ denoting the relative permittivity of the IL layer. The characteristics dependent on the thickness of the FE material are given as \cite{lyu_thickness}
\begin{equation}
    \mathrm{P_{s,i}=h_{s}t_{fe,i} , \quad P_{r,i} = h_{r}t_{fe,i} , \quad E_{c,i} = h_{c}t_{fe,i}^{-0.61}}
    \label{thickness_preisach}
\end{equation}
where, $\mathrm{P_{s,i}}$, $\mathrm{P_{r,i}}$, and $\mathrm{E_{c,i}}$ are the FEs saturated polarization, remanent polarization, and coercive field, respectively. $\mathrm{P_{r,i}}$ and $\mathrm{P_{s,i}}$ are proportional to FE thickness $\mathrm{t_{fe,i}}$, and $\mathrm{E_{c,i}}$ are inversely proportional to $\mathrm{t_{fe,i}}$ following a slope of -0.61, which is in agreement with the scaling value of $-2/3$.\cite{janovec1958theory_ratio} The proportionality factors in relation of $\mathrm{P_{s,i}}$, $\mathrm{P_{r,i}}$, and $\mathrm{E_{c,i}}$ are represented by the constants $\mathrm{h_{s}}$, $\mathrm{h_{r}}$ and $\mathrm{h_{c}}$, respectively. FE polarization charge density $\mathrm{P_{fe,i}}$ is given by the Preisach-based model, \cite{miller1990device,sutor2010preisach,mayergoyz1988generalized} 
\begin{equation}
    \mathrm{P_{fe,i}=P_{s,i} \tanh \left[\ln \left(\frac{P_{s,i} + P_{r,i}}{P_{s,i}-P_{r,i}}\right) \cdot \frac{V_{fe,i} \pm V_{c,i}}{2 V_{c,i}}\right]}
    \label{preisach_model}
\end{equation}
\begin{figure}[htb!]
    \centering
    \includegraphics[scale=0.58]{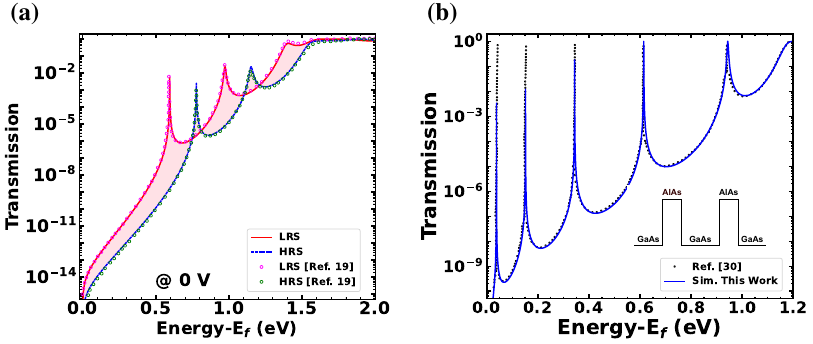}
    \caption{Comparison of simulated (a) and (b) Transmission-Energy curve with references.~\cite{chang2023edl,ando1987calculation}}
    \label{3reference}
\end{figure}
where $\mathrm{V_{c,i}}$ is the coercive voltage. The calculation of the self-consistent current for multilayer FTJ employs the integration of a Preisach-based model with the NEGF formalism, as depicted in \hyperref[2simframe]{Fig.~\ref{2simframe}}. The self-consistent solution of Eq.~\hyperref[thomas_fermi]{(\ref{thomas_fermi})} to \hyperref[preisach_model]{(\ref{preisach_model})} determines the electrostatic potential, FE polarization, and surface charge density for each $\mathrm{V_a}$ value. This process continues iteratively until the difference between the updated and initial $\mathrm{\sigma_s}$ reaches satisfactory convergence. The NEGF technique, suitable for quantum transport phenomena, evaluates the transmission coefficient $\mathrm{T_m}$ and current density J following the calculation of the energy band profiles. The Landauer formula is subsequently utilized to calculate the current density, which is defined as\cite{zhou2021time,datta2005quantum,sharma2016ultrasensitive}
\begin{equation}
    \mathrm{J=-\sum_{k_y,k_z}^{} \frac{2 e}{A h} \int T_m(E) \left[f_{1}(E)-f_{2}(E)\right]dE }
    \label{negf_current}
\end{equation}
where, $\mathrm{k_{y}}$ and $\mathrm{k_{z}}$ represent the electron wave vectors within the transverse plane, while $\mathrm{f_1}$ and $\mathrm{f_2}$ denote the Fermi-Dirac distribution functions associated with the TE and BE, respectively. Further, \textquotesingle e\textquotesingle~
signifies the electron charge, and \textquotesingle h\textquotesingle~ refers to Planck’s constant. \hyperref[3reference]{Figure~\ref{3reference}\textcolor{blue}{(a)}} shows that the simulation approaches were thoroughly examined by comparing the simulated transmission-energy curve of a double-barrier structure with a single QW (HZO/$\mathrm{Ta_2 O_5}$/HZO) with existing data. \cite{chang2023edl} This comparison showed perfect alignment, confirming model accuracy. The transmission-energy curve for a structure (AlAs/GaAs/AlAs) with a single QW, where RT is observed, was replicated and agreed with simulated results [\hyperref[3reference]{Fig.~\ref{3reference}\textcolor{blue}{(b)}}].\cite{ando1987calculation} The excellent agreement between the simulated and reference results highlights the robustness and reliability of the approach. 
\begin{figure}[htb!]
    \centering
    \includegraphics[scale=0.58]{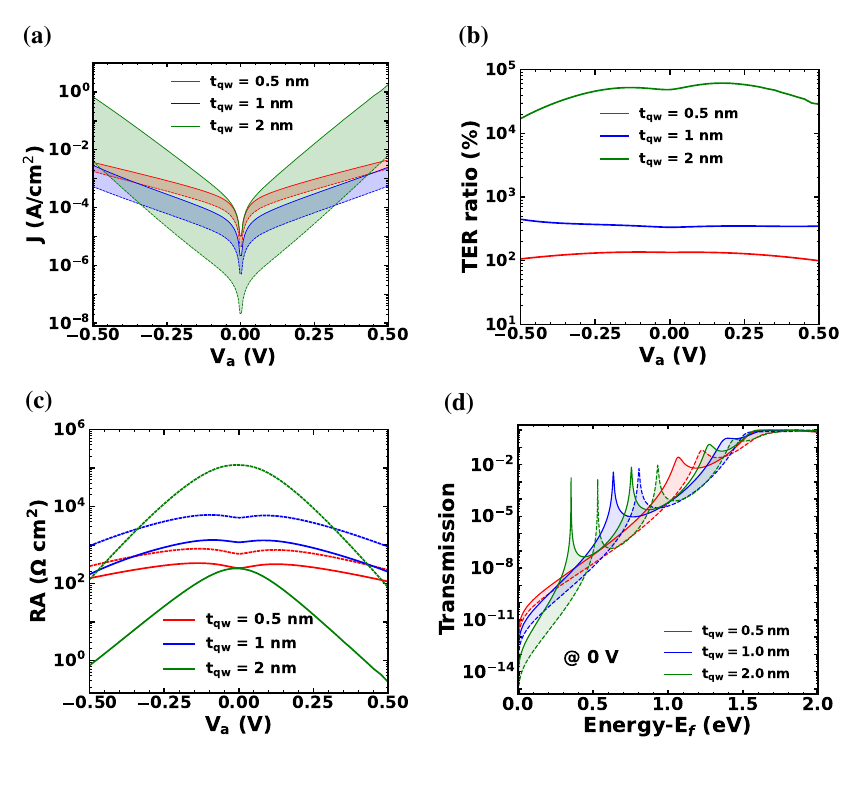}
    \caption{(a) Read current density-voltage (J-V) curves for the $\mathrm{TiN/HZO/TiO_2/HZO/TiN}$ (MFIFM) stack for $\mathrm{V_a}$=$\pm 0.5$V. Corresponding Fig.~(b) shows the TER ratio ($\%$), (c) RA product, and (d) transmission characteristics (0V) at different values of thickness $\mathrm{t_{qw}}$. Here, in Fig.~(a), (c), and (d), bold and dashed lines indicate the LRS and HRS, respectively.}
    \label{4pm_p5v}
\end{figure}

In the simulation framework, we utilize the following parameter values: The electron affinity ${\mathrm{\chi}}$ is set at $2.8$ eV for $\mathrm{HZO}$\cite{fan2016ferroelectricity_hzo} and $4.0$ eV for $\mathrm{TiO_2}$,\cite{robertson2013band} and the metal TiN is characterized by a work function $\Phi_{\mathrm{m}}$\cite{fillot2005investigations} of $4.3$ eV with a screening length~\cite{chang2023ted} ($\mathrm{\lambda_{m1}=\lambda_{m2}}$) of $0.5$$\text{\AA}$. The materials  $\mathrm{HZO}$~\cite{materlik2015origin}, $\mathrm{TiO_2}$~\cite{mondal2016tunable} and TiN\cite{tripura2014investigation} exhibited relative permittivities of $25$, $48$, and $3$, respectively. The effective electron masses are considered as $0.14$$\mathrm{m_o}$ for $\mathrm{HZO}$ \cite{monaghan2009determination}  and $0.5$$\mathrm{m_o}$ for $\mathrm{TiO_2}$.\cite{gulomov2023investigation} The thickness of the $\mathrm{HZO}$, is defined at $4$nm, exhibiting a spontaneous polarization of $\mathrm{P_s}=5.5$$\mathrm{\mu C/cm^2}$, a remanent polarization of $\mathrm{P_r}=5$$\mathrm{\mu C/cm^2}$, and a coercive field of $\mathrm{E_c}=1$$\mathrm{MV/cm}$. \cite{chang2023ted} The thicknesses of the FE layers, denoted as $\mathrm{t_{fe1}}$ and $\mathrm{t_{fe2}}$, have been determined at $3$ and $1$ nm, respectively. The parameters $\mathrm{h_s}$, $\mathrm{h_r}$, and $\mathrm{h_c}$ can derive using these values as standards [Eq.~\hyperref[thickness_preisach]{(\ref{thickness_preisach})}]. \\
We begin with investigating the impact of different QW thicknesses on the performance of the MFIFM FTJ. This investigation governs an essential stage in understanding the ways that structural parameters affect device characteristics.
\hyperref[4pm_p5v]{Figure~\ref{4pm_p5v}\textcolor{blue}{(a)}} illustrates current density-voltage (J-V) characteristics for both LRS and HRS of the $\mathrm{TiN/HZO/TiO_2/HZO/TiN}$ stack, with $\mathrm{V_a}$ kept at $\pm0.5$V across QW thicknesses of 0.5, 1, and 2 nm. A notable distinction in the current density has been observed between LRS and HRS when the QW thickness is set at 2 nm. As the thickness of the well $\mathrm{t_{qw}}$ increases in a single QW FTJ, the tunneling currents are affected by the interaction between direct and resonant tunneling. Initially, increasing $\mathrm{t_{qw}}$ (at 1nm) widens the barrier, reducing direct tunneling and thereby lowering the current density in both LRS and HRS. However, as the quantized energy levels shift closer to the Fermi level, $\mathrm{E_{f}}$, RT is enhanced, causing the higher current density in LRS. 
% Beyond a certain $\mathrm{t_{qw}}$, further thickening moves these energy levels away from optimal alignment with $\mathrm{E_{F}}$, reducing RT efficiency and leading to a subsequent decline in LRS. At the same time, HRS is consistently declining as a result of the inhibition of direct tunneling. 
This leads to a significant boost in the TER ratio, improving the performance of the device. Furthermore, with an increase in current density in LRS attributed to enhanced RT, there is a reduction in the RA product, leading to improved readability and power efficiency.

\hyperref[4pm_p5v]{Figs.~\ref{4pm_p5v}\textcolor{blue}{(b)}} and \hyperref[4pm_p5v]{\ref{4pm_p5v}\textcolor{blue}{(c)}} illustrate the characteristics of TER ratio and the RA product, respectively. We have taken 0.15V applied voltage as a reference point to compare these characteristics. At $\mathrm{t_{qw}}$=0.5 nm, the TER ratio and RA product are $1.34 \times 10^{2}$\% and $3.12 \times 10^{2}$  $\Omega$-$\text{cm}^2$, respectively, and at
$\mathrm{t_{qw}} $=1nm, we found the TER ratio and RA product are approximately $3.48 \times 10^{2}$\% and $1.20 \times 10^{3}$ $ \Omega$-$\text{cm}^2$, respectively. The TER ratio and RA product are found to be approximately $6.11\times 10^{4}$\% and $6.68 \times 10^{1}$ $\Omega$-$\text{cm}^2$, respectively, for $\mathrm{t_{qw}}$ =2nm. Consequently, at $\mathrm{t_{qw}} = 2$nm, the TER ratio is enhanced by approximately $1.75 \times 10^2$ times, while the RA product is reduced by nearly $5\times 10^{-2}$ times compared to a QW thickness of 1 nm. As shown in \hyperref[4pm_p5v]{Fig.~\ref{4pm_p5v}\textcolor{blue}{(d)}}, $\mathrm{t_{qw}}$ increases from 0.5 to 2 nm, the transmission characteristics undergo notable changes. At $\mathrm{t_{qw}}$=0.5 nm, the QW is extremely narrow, supporting only a single resonant state, which results in a broad transmission peak at higher energy levels. The strong coupling between the barriers leads to a relatively wide peak with moderate transmission. As the QW width increases to 1 nm, additional quantized states form within the well, creating multiple sharper transmission peaks. These peaks shift to lower energy levels as the confined energy levels decrease with the increasing well width. Consequently, the transmission probability at resonant energies also increases, enabling more efficient electron tunneling. At 2 nm, the well accommodates even more resonant states, causing the transmission peaks to become denser and more distinct. The energy spacing between the peaks continues to decrease as quantum confinement weakens. This results in higher overall transmission, as more electrons can tunnel resonantly through the structure. 
\begin{figure}[htb!]
    \centering
    \includegraphics[scale=0.58]{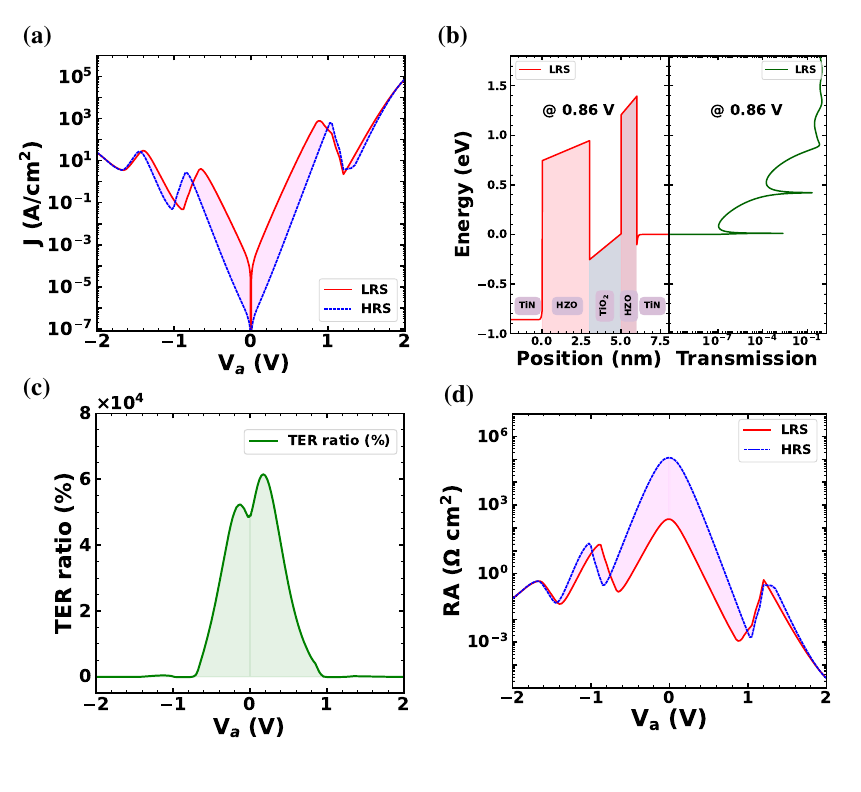}
    \caption{(a) Current density-voltage (J-V) characteristics with sweep voltage $\pm$2V of MFIFM  stack ($\mathrm{TiN/HZO/TiO_2/HZO/TiN}$) FTJ at $\mathrm{t_{qw}} = 2\mathrm{nm}$. (b) Band diagram and transmission curve when the current peak takes place at 0.86V. (c) TER ratio (\%) and (d)  RA product characteristics.}
    \label{5iv_2v}
\end{figure}

Next, we analyze the J-V characteristics over a voltage sweep range of $\pm 2$V with $\mathrm{t_{qw}}$=2 $\mathrm{nm}$, as illustrated in \hyperref[5iv_2v]{Fig.~\ref{5iv_2v}\textcolor{blue}{(a)}}. The band diagram and transmission at 0.86V, where the current peak occurs, are illustrated in \hyperref[5iv_2v]{Fig.~\ref{5iv_2v}\textcolor{blue}{(b)}}. The resonant peak near 0.86V highlights the dominance of RT, generated by the alignment of the resonant peak with the Fermi level, while the negative differential resistance (NDR) effect is observed at this voltage. This illustrates that when the Fermi energy $\mathrm{E_{f}}$ in the electrode aligns with the quantized energy levels of the structure, the current reaches a maximum. \hyperref[5iv_2v]{Figs.~\ref{5iv_2v}\textcolor{blue}{(c)}} and {\ref{5iv_2v}\textcolor{blue}{(d)}} depict the TER ratio and RA product for the multi-barrier stack FTJ. While TER ratio shows a significant enhancement at lower bias voltages, peaking approximately $6.15 \times 10^{4}$\% at 0.175V with the low RA product is measured to be $4.71 \times 10^{1}$~$\Omega$-$\mathrm{cm}^2$ enabling better readout and cross-section scaling for high density.\cite{abuwasib2016scaling}
\begin{figure}[htb!]
    \centering
    \includegraphics[scale=0.58]{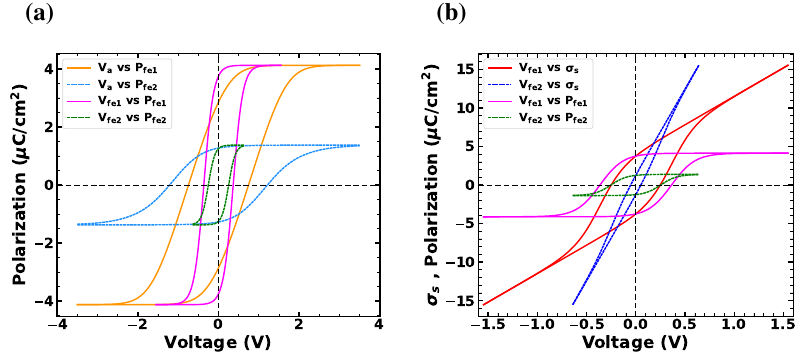}
    \caption{Hysteresis characteristics of $\mathrm{TiN/HZO/TiO_2/HZO/TiN}$ (MFIFM) stack at QW thickness $\mathrm{t_{qw}}$=2$\mathrm{nm}$. (a) Polarization–Voltage (P–V) curves showing FE switching behavior. (b) Polarization and surface charge density ($\mathrm{\sigma_s}$) as functions of voltage.}
    \label{hyst}
\end{figure}
%\hyperref[5iv_2v]{Fig.~\ref{5iv_2v}\textcolor{blue}{(a)}} shows that the proposed FTJ design exhibits a valley of NDR that nearly coincides with the switching voltage, enabling energy-efficient operation with minimal power loss (power loss = $\mathrm{I\times V}$).

Following this, \hyperref[hyst]{Fig. \ref{hyst}\textcolor{blue}{}} shows the hysteresis behavior of the MFIFM structure at $\mathrm{t_{qw}} = 2\mathrm{nm}$. \hyperref[hyst]{Fig. \ref{hyst}\textcolor{blue}{(a)}} presents the $\mathrm{V_{a}}$–$\mathrm{P_{fe,i}}$ hysteresis loops, along with the corresponding FE voltages ($\mathrm{V_{fe,i}}$) under an applied sweep. \hyperref[hyst]{Fig.~\ref{hyst}\textcolor{blue}{(b)}} illustrates the dependence of polarization and $\mathrm{\sigma_s}$ on the FE voltages. The J–V characteristics of the MFIFM structure reflect polarization switching in the FE layers [\hyperref[5iv_2v]{Fig.~\ref{5iv_2v}\textcolor{blue}{(a)}}]. Around $V_a = \pm 1.6$V, the polarization difference is minimal, indicating aligned polarization states. At lower voltages, a mismatch in polarization creates an internal field imbalance, resulting in distinct LRS and HRS. While the polarization hysteresis remains symmetric, the unequal thicknesses of the FE layers introduce structural asymmetry, resulting in an asymmetric 
J–V loop.

% Following this, we analyze the hysteresis characteristics of the MFIFM structure at a QW thickness of $\mathrm{t_{qw}} = 2 \mathrm{nm}$, as shown in \hyperref[hyst]{Fig.~\ref{hyst}\textcolor{blue}{}}. The hysteresis curves represent the polarization switching of two FE layers ($\mathrm{P_{fe1}}$ \& $\mathrm{P_{fe2}}$) under an applied voltage sweep, as shown in \hyperref[hyst]{Fig.~\ref{hyst}\textcolor{blue}{(a)}}. The figure also depicts the corresponding voltages across each FE layer, $\mathrm{V_{fe1}}$ and $\mathrm{V_{fe2}}$, highlighting their switching characteristics in response to the applied voltage.\hyperref[hyst]{Fig.~\ref{hyst}\textcolor{blue}{(b)}} illustrates the dependence of polarization and surface charge density ($\mathrm{\sigma_s}$) on the FE voltages, highlighting how polarization modulates the surface charge density within the structure.

% As the applied voltage increases, one of the FE layers (typically $\mathrm{P_{fe1}}$) switches polarization first, leading to a significant redistribution of the internal electric field across the structure. This switching can temporarily reduce the effective field across the tunneling region, thereby lowering the current despite increasing voltage — the hallmark of NDR. When the second layer ($\mathrm{P_{fe2}}$) begins to switch, the internal field can realign again, restoring or even enhancing current flow. This process creates a voltage window where the current decreases with increasing voltage, giving rise to NDR. The hysteresis in polarization switching explains why the NDR appears asymmetrically and depends on the sweep direction.

\begin{figure}[htb!]
    \centering
    \includegraphics[scale=0.6]{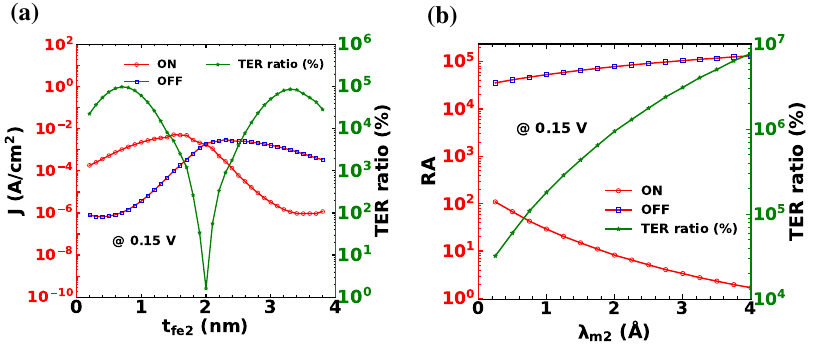}
    \caption{ Impact on the performance of multilayer
    MFIFM FTJ due to (a) QW position with varying $\mathrm{t_{fe2}=(t_{fe1}-4) nm}$ at read voltage $\mathrm{V_{r}}$ = 0.15V and (b) screening length with varying $\mathrm{\lambda_{m2}}$, and $\mathrm{\lambda_{m1}}$ is fixed at 0.5 $\text{\AA}$ at $\mathrm{V_{r}}$ = 0.15V.}
    \label{6tfe2_lambda2}
\end{figure}

\hyperref[6tfe2_lambda2]{Figure~\ref{6tfe2_lambda2}\textcolor{blue}{(a)}} illustrates the effect of the QW position ($\mathrm{t_{qw}}$=2 $\mathrm{nm}$) while varying the thickness of the second FE layer defined as $\mathrm{t_{fe2}=(t_{fe1} - 4) nm}$. The read voltage $\mathrm{V_{r}}$ is set to 0.15V, a value that is sufficiently low to ensure that the polarization stays constant throughout the read process. It is observed that as the thickness $\mathrm{t_{fe2}}$ increases, the ON state current exhibits an initial increase, followed by a decrease to approximately zero once the thickness reaches 2 nm. The observed behavior can be linked to the asymmetry of the barrier, which arises from the differences in $\mathrm{t_{fe1}}$ and $\mathrm{t_{fe2}}$. The changes noted in both the ON and OFF states (LRS and HRS) are a direct result of this asymmetry. \hyperref[6tfe2_lambda2]
{Fig.~\ref{6tfe2_lambda2}\textcolor{blue}{(b)}} illustrates the effect of the screening length, particularly with varying values of $\mathrm{\lambda_{m2}}$, while maintaining $\mathrm{\lambda_{m1}}$ constant for the MFIFM FTJ. Differences in screening length characteristics lead to a notable increase in the TER ratio at a read voltage of 0.15\,V. The resulting asymmetry in interfacial screening further enhances the TER ratio, highlighting the vital role of material properties in achieving maximum device efficiency. 

In conclusion, we have examined a multilayer FTJ design that uses a QW between two FE layers in place of the conventional metal-ferroelectric-metal arrangement. Through a self-consistent coupling of NEGF with the Preisach-based model, we show that the QW enhances resonance effects, resulting in a notable increase in the TER ratio and a notable drop in the RA product. The NDR effect enhances the functionality of the device. The potential applications are thus in logic components, programmable resistors, and improved non-volatile memory. Future research in FE materials and QW characteristics could thus lead to more effective, scalable devices for neuromorphic computing and next-generation electronics.

\bibliography{cite_link}

%merlin.mbs aipnum4-1.bst 2010-07-25 4.21a (PWD, AO, DPC) hacked
%Control: key (0)
%Control: author (8) initials jnrlst
%Control: editor formatted (1) identically to author
%Control: production of article title (0) allowed
%Control: page (1) range
%Control: year (1) truncated
%Control: production of eprint (0) enabled
\begin{thebibliography}{43}%
\makeatletter
\providecommand \@ifxundefined [1]{%
 \@ifx{#1\undefined}
}%
\providecommand \@ifnum [1]{%
 \ifnum #1\expandafter \@firstoftwo
 \else \expandafter \@secondoftwo
 \fi
}%
\providecommand \@ifx [1]{%
 \ifx #1\expandafter \@firstoftwo
 \else \expandafter \@secondoftwo
 \fi
}%
\providecommand \natexlab [1]{#1}%
\providecommand \enquote  [1]{``#1''}%
\providecommand \bibnamefont  [1]{#1}%
\providecommand \bibfnamefont [1]{#1}%
\providecommand \citenamefont [1]{#1}%
\providecommand \href@noop [0]{\@secondoftwo}%
\providecommand \href [0]{\begingroup \@sanitize@url \@href}%
\providecommand \@href[1]{\@@startlink{#1}\@@href}%
\providecommand \@@href[1]{\endgroup#1\@@endlink}%
\providecommand \@sanitize@url [0]{\catcode `\\12\catcode `\$12\catcode `\&12\catcode `\#12\catcode `\^12\catcode `\_12\catcode `\%12\relax}%
\providecommand \@@startlink[1]{}%
\providecommand \@@endlink[0]{}%
\providecommand \url  [0]{\begingroup\@sanitize@url \@url }%
\providecommand \@url [1]{\endgroup\@href {#1}{\urlprefix }}%
\providecommand \urlprefix  [0]{URL }%
\providecommand \Eprint [0]{\href }%
\providecommand \doibase [0]{http://dx.doi.org/}%
\providecommand \selectlanguage [0]{\@gobble}%
\providecommand \bibinfo  [0]{\@secondoftwo}%
\providecommand \bibfield  [0]{\@secondoftwo}%
\providecommand \translation [1]{[#1]}%
\providecommand \BibitemOpen [0]{}%
\providecommand \bibitemStop [0]{}%
\providecommand \bibitemNoStop [0]{.\EOS\space}%
\providecommand \EOS [0]{\spacefactor3000\relax}%
\providecommand \BibitemShut  [1]{\csname bibitem#1\endcsname}%
\let\auto@bib@innerbib\@empty
%</preamble>
\bibitem [{\citenamefont {Park}\ \emph {et~al.}(2024)\citenamefont {Park}, \citenamefont {Lee}, \citenamefont {Park}, \citenamefont {Kim},\ and\ \citenamefont {Jang}}]{park2024ferroelectric}%
  \BibitemOpen
  \bibfield  {author} {\bibinfo {author} {\bibfnamefont {S.~H.}\ \bibnamefont {Park}}, \bibinfo {author} {\bibfnamefont {H.~J.}\ \bibnamefont {Lee}}, \bibinfo {author} {\bibfnamefont {M.~H.}\ \bibnamefont {Park}}, \bibinfo {author} {\bibfnamefont {J.}~\bibnamefont {Kim}}, \ and\ \bibinfo {author} {\bibfnamefont {H.~W.}\ \bibnamefont {Jang}},\ }\href@noop {} {\bibfield  {journal} {\bibinfo  {journal} {\href{https://doi.org/10.1088/1361-6463/ad33f5}{\textcolor{blue}{J. Phys. D: Appl. Phys. }}}\ }\textbf {\bibinfo {volume} {\textbf{57}\normalfont(25)}},\ \bibinfo {pages} {253002} (\bibinfo {year} {2024})}\BibitemShut {NoStop}%
\bibitem [{\citenamefont {Kohlstedt}\ \emph {et~al.}(2005)\citenamefont {Kohlstedt}, \citenamefont {Pertsev}, \citenamefont {Rodr{\'\i}guez~Contreras},\ and\ \citenamefont {Waser}}]{Kohlstedt2005}%
  \BibitemOpen
  \bibfield  {author} {\bibinfo {author} {\bibfnamefont {H.}~\bibnamefont {Kohlstedt}}, \bibinfo {author} {\bibfnamefont {N.}~\bibnamefont {Pertsev}}, \bibinfo {author} {\bibfnamefont {J.}~\bibnamefont {Rodr{\'\i}guez~Contreras}}, \ and\ \bibinfo {author} {\bibfnamefont {R.}~\bibnamefont {Waser}},\ }\href@noop {} {\bibfield  {journal} {\bibinfo  {journal} {\href{https://doi.org/10.1103/PhysRevB.72.125341}{\textcolor{blue}{Phys. Rev. B}}}\ }\textbf {\bibinfo {volume} {\textbf{72}\normalfont(12)}},\ \bibinfo {pages} {125341} (\bibinfo {year} {2005})}\BibitemShut {NoStop}%
\bibitem [{\citenamefont {Luo}\ \emph {et~al.}(2025)\citenamefont {Luo}, \citenamefont {Ma}, \citenamefont {Sando}, \citenamefont {Zhang},\ and\ \citenamefont {Valanoor}}]{luo2025review}%
  \BibitemOpen
  \bibfield  {author} {\bibinfo {author} {\bibfnamefont {K.-F.}\ \bibnamefont {Luo}}, \bibinfo {author} {\bibfnamefont {Z.}~\bibnamefont {Ma}}, \bibinfo {author} {\bibfnamefont {D.}~\bibnamefont {Sando}}, \bibinfo {author} {\bibfnamefont {Q.}~\bibnamefont {Zhang}}, \ and\ \bibinfo {author} {\bibfnamefont {N.}~\bibnamefont {Valanoor}},\ }\href@noop {} {\bibfield  {journal} {\bibinfo  {journal} {\href{https://doi.org/10.1021/acsnano.4c14446}{\textcolor{blue} {ACS nano}}}\ }\textbf {\bibinfo {volume} {\textbf{19}\normalfont(7)}},\ \bibinfo {pages} {6622--6647} (\bibinfo {year} {2025})}\BibitemShut {NoStop}%
\bibitem [{\citenamefont {Kobayashi}\ \emph {et~al.}(2019)\citenamefont {Kobayashi}, \citenamefont {Tagawa}, \citenamefont {Mo}, \citenamefont {Saraya},\ and\ \citenamefont {Hiramoto}}]{Kobayashi2019}%
  \BibitemOpen
  \bibfield  {author} {\bibinfo {author} {\bibfnamefont {M.}~\bibnamefont {Kobayashi}}, \bibinfo {author} {\bibfnamefont {Y.}~\bibnamefont {Tagawa}}, \bibinfo {author} {\bibfnamefont {F.}~\bibnamefont {Mo}}, \bibinfo {author} {\bibfnamefont {T.}~\bibnamefont {Saraya}}, \ and\ \bibinfo {author} {\bibfnamefont {T.}~\bibnamefont {Hiramoto}},\ }\href@noop {} {\bibfield  {journal} {\bibinfo  {journal} {\href{https://doi.org/10.1109/JEDS.2018.2885932}{\textcolor{blue}{IEEE J. Electron Devices Soc.}}}\ }\textbf {\bibinfo {volume} {7}},\ \bibinfo {pages} {134--139} (\bibinfo {year} {2019})}\BibitemShut {NoStop}%
\bibitem [{\citenamefont {Covi}\ \emph {et~al.}(2022)\citenamefont {Covi}, \citenamefont {Mulaosmanovic}, \citenamefont {Max}, \citenamefont {Slesazeck},\ and\ \citenamefont {Mikolajick}}]{covi2022ferroelectric}%
  \BibitemOpen
  \bibfield  {author} {\bibinfo {author} {\bibfnamefont {E.}~\bibnamefont {Covi}}, \bibinfo {author} {\bibfnamefont {H.}~\bibnamefont {Mulaosmanovic}}, \bibinfo {author} {\bibfnamefont {B.}~\bibnamefont {Max}}, \bibinfo {author} {\bibfnamefont {S.}~\bibnamefont {Slesazeck}}, \ and\ \bibinfo {author} {\bibfnamefont {T.}~\bibnamefont {Mikolajick}},\ }\href@noop {} {\bibfield  {journal} {\bibinfo  {journal} {\href{https://doi.org/10.1088/2634-4386/ac4918}{\textcolor{blue}{Neuromorphic Comput. and Eng.}}}\ }\textbf {\bibinfo {volume} {\textbf{2}\normalfont(1)}},\ \bibinfo {pages} {012002} (\bibinfo {year} {2022})}\BibitemShut {NoStop}%
\bibitem [{\citenamefont {Hwang}, \citenamefont {Goh},\ and\ \citenamefont {Jeon}(2024)}]{hwang2024smallreview}%
  \BibitemOpen
  \bibfield  {author} {\bibinfo {author} {\bibfnamefont {J.}~\bibnamefont {Hwang}}, \bibinfo {author} {\bibfnamefont {Y.}~\bibnamefont {Goh}}, \ and\ \bibinfo {author} {\bibfnamefont {S.}~\bibnamefont {Jeon}},\ }\href@noop {} {\bibfield  {journal} {\bibinfo  {journal} {\href{https://doi.org/10.1002/smll.202305271}{\textcolor{blue} {Small}}}\ }\textbf {\bibinfo {volume} {\textbf{20}\normalfont(9)}},\ \bibinfo {pages} {2305271} (\bibinfo {year} {2024})}\BibitemShut {NoStop}%
\bibitem [{\citenamefont {Jao}\ \emph {et~al.}(2021)\citenamefont {Jao}, \citenamefont {Xiao}, \citenamefont {Saha}, \citenamefont {Gupta},\ and\ \citenamefont {Narayanan}}]{jao2021design}%
  \BibitemOpen
  \bibfield  {author} {\bibinfo {author} {\bibfnamefont {N.}~\bibnamefont {Jao}}, \bibinfo {author} {\bibfnamefont {Y.}~\bibnamefont {Xiao}}, \bibinfo {author} {\bibfnamefont {A.~K.}\ \bibnamefont {Saha}}, \bibinfo {author} {\bibfnamefont {S.~K.}\ \bibnamefont {Gupta}}, \ and\ \bibinfo {author} {\bibfnamefont {V.}~\bibnamefont {Narayanan}},\ }\href@noop {} {\bibfield  {journal} {\bibinfo  {journal} {\href{http://dx.doi.org/10.1109/JXCDC.2021.3117566}{\textcolor{blue}{IEEE J. Explor. Solid-State Comput. Devices and Circuits}}}\ }\textbf {\bibinfo {volume} {\textbf{7}\normalfont(2)}},\ \bibinfo {pages} {115--122} (\bibinfo {year} {2021})}\BibitemShut {NoStop}%
\bibitem [{\citenamefont {Goh}\ \emph {et~al.}(2020)\citenamefont {Goh}, \citenamefont {Hwang}, \citenamefont {Lee}, \citenamefont {Kim},\ and\ \citenamefont {Jeon}}]{goh2020ultra}%
  \BibitemOpen
  \bibfield  {author} {\bibinfo {author} {\bibfnamefont {Y.}~\bibnamefont {Goh}}, \bibinfo {author} {\bibfnamefont {J.}~\bibnamefont {Hwang}}, \bibinfo {author} {\bibfnamefont {Y.}~\bibnamefont {Lee}}, \bibinfo {author} {\bibfnamefont {M.}~\bibnamefont {Kim}}, \ and\ \bibinfo {author} {\bibfnamefont {S.}~\bibnamefont {Jeon}},\ }\href@noop {} {\bibfield  {journal} {\bibinfo  {journal} {\href{https://doi.org/10.1063/5.0029516}{\textcolor{blue}{Appl. Phys. Lett.}}}\ }\textbf {\bibinfo {volume} {\textbf{117}\normalfont(24)}},\ \bibinfo {pages} {242901} (\bibinfo {year} {2020})}\BibitemShut {NoStop}%
\bibitem [{\citenamefont {Ambriz-Vargas}\ \emph {et~al.}(2017)\citenamefont {Ambriz-Vargas}, \citenamefont {Kolhatkar}, \citenamefont {Thomas}, \citenamefont {Nouar}, \citenamefont {Sarkissian}, \citenamefont {Gomez-Y{\'a}{\~n}ez}, \citenamefont {Gauthier},\ and\ \citenamefont {Ruediger}}]{ambriz2017tunneling}%
  \BibitemOpen
  \bibfield  {author} {\bibinfo {author} {\bibfnamefont {F.}~\bibnamefont {Ambriz-Vargas}}, \bibinfo {author} {\bibfnamefont {G.}~\bibnamefont {Kolhatkar}}, \bibinfo {author} {\bibfnamefont {R.}~\bibnamefont {Thomas}}, \bibinfo {author} {\bibfnamefont {R.}~\bibnamefont {Nouar}}, \bibinfo {author} {\bibfnamefont {A.}~\bibnamefont {Sarkissian}}, \bibinfo {author} {\bibfnamefont {C.}~\bibnamefont {Gomez-Y{\'a}{\~n}ez}}, \bibinfo {author} {\bibfnamefont {M.}~\bibnamefont {Gauthier}}, \ and\ \bibinfo {author} {\bibfnamefont {A.}~\bibnamefont {Ruediger}},\ }\href@noop {} {\bibfield  {journal} {\bibinfo  {journal} {\href{https://doi.org/10.1063/1.4977028}{\textcolor{blue}{Appl. Phys. Lett.}}}\ }\textbf {\bibinfo {volume} {\textbf{110}\normalfont(9)}},\ \bibinfo {pages} {093106} (\bibinfo {year} {2017})}\BibitemShut {NoStop}%
\bibitem [{\citenamefont {Max}\ \emph {et~al.}(2020)\citenamefont {Max}, \citenamefont {Hoffmann}, \citenamefont {Mulaosmanovic}, \citenamefont {Slesazeck},\ and\ \citenamefont {Mikolajick}}]{max2020hafnia}%
  \BibitemOpen
  \bibfield  {author} {\bibinfo {author} {\bibfnamefont {B.}~\bibnamefont {Max}}, \bibinfo {author} {\bibfnamefont {M.}~\bibnamefont {Hoffmann}}, \bibinfo {author} {\bibfnamefont {H.}~\bibnamefont {Mulaosmanovic}}, \bibinfo {author} {\bibfnamefont {S.}~\bibnamefont {Slesazeck}}, \ and\ \bibinfo {author} {\bibfnamefont {T.}~\bibnamefont {Mikolajick}},\ }\href@noop {} {\bibfield  {journal} {\bibinfo  {journal} {\href{https://doi.org/10.1021/acsaelm.0c00832}{\textcolor{blue}{ ACS Appl. Electron. Mater.}}}\ }\textbf {\bibinfo {volume} {\textbf{2}\normalfont(12)}},\ \bibinfo {pages} {4023--4033} (\bibinfo {year} {2020})}\BibitemShut {NoStop}%
\bibitem [{\citenamefont {Athle}\ and\ \citenamefont {Borg}(2024)}]{athle2024ferroelectric}%
  \BibitemOpen
  \bibfield  {author} {\bibinfo {author} {\bibfnamefont {R.}~\bibnamefont {Athle}}\ and\ \bibinfo {author} {\bibfnamefont {M.}~\bibnamefont {Borg}},\ }\href@noop {} {\bibfield  {journal} {\bibinfo  {journal} {\href{https://doi.org/10.1002/aisy.202300554}{\textcolor{blue}{Adv.Intell. Syst. }}}\ }\textbf {\bibinfo {volume} {\textbf{6}\normalfont(3)}},\ \bibinfo {pages} {2300554} (\bibinfo {year} {2024})}\BibitemShut {NoStop}%
\bibitem [{\citenamefont {Yoon}\ \emph {et~al.}(2019)\citenamefont {Yoon}, \citenamefont {Hong}, \citenamefont {Song}, \citenamefont {Ahn},\ and\ \citenamefont {Ahn}}]{Yoon2019}%
  \BibitemOpen
  \bibfield  {author} {\bibinfo {author} {\bibfnamefont {J.}~\bibnamefont {Yoon}}, \bibinfo {author} {\bibfnamefont {S.}~\bibnamefont {Hong}}, \bibinfo {author} {\bibfnamefont {Y.~W.}\ \bibnamefont {Song}}, \bibinfo {author} {\bibfnamefont {J.-H.}\ \bibnamefont {Ahn}}, \ and\ \bibinfo {author} {\bibfnamefont {S.-E.}\ \bibnamefont {Ahn}},\ }\href@noop {} {\bibfield  {journal} {\bibinfo  {journal} {\href{https://doi.org/10.1063/1.5119948}{\textcolor{blue}{Appl. Phys. Lett.}}}\ }\textbf {\bibinfo {volume} {\textbf{115}\normalfont(15)}},\ \bibinfo {pages} {153502} (\bibinfo {year} {2019})}\BibitemShut {NoStop}%
\bibitem [{\citenamefont {Wang}\ \emph {et~al.}(2016)\citenamefont {Wang}, \citenamefont {Cho}, \citenamefont {Shin}, \citenamefont {Kim}, \citenamefont {Das}, \citenamefont {Yoon}, \citenamefont {Chung},\ and\ \citenamefont {Noh}}]{wang2016overcoming}%
  \BibitemOpen
  \bibfield  {author} {\bibinfo {author} {\bibfnamefont {L.}~\bibnamefont {Wang}}, \bibinfo {author} {\bibfnamefont {M.~R.}\ \bibnamefont {Cho}}, \bibinfo {author} {\bibfnamefont {Y.~J.}\ \bibnamefont {Shin}}, \bibinfo {author} {\bibfnamefont {J.~R.}\ \bibnamefont {Kim}}, \bibinfo {author} {\bibfnamefont {S.}~\bibnamefont {Das}}, \bibinfo {author} {\bibfnamefont {J.-G.}\ \bibnamefont {Yoon}}, \bibinfo {author} {\bibfnamefont {J.-S.}\ \bibnamefont {Chung}}, \ and\ \bibinfo {author} {\bibfnamefont {T.~W.}\ \bibnamefont {Noh}},\ }\href@noop {} {\bibfield  {journal} {\bibinfo  {journal} {\href{http://dx.doi.org/10.1021/acs.nanolett.6b01418}{\textcolor{blue}{Nano letters}}}\ }\textbf {\bibinfo {volume} {\textbf{16}\normalfont(6)}},\ \bibinfo {pages} {3911--3918} (\bibinfo {year} {2016})}\BibitemShut {NoStop}%
\bibitem [{\citenamefont {Lu}\ \emph {et~al.}(2014)\citenamefont {Lu}, \citenamefont {Lipatov}, \citenamefont {Ryu}, \citenamefont {Kim}, \citenamefont {Lee}, \citenamefont {Zhuravlev}, \citenamefont {Eom}, \citenamefont {Tsymbal}, \citenamefont {Sinitskii},\ and\ \citenamefont {Gruverman}}]{lu2014ferroelectric}%
  \BibitemOpen
  \bibfield  {author} {\bibinfo {author} {\bibfnamefont {H.}~\bibnamefont {Lu}}, \bibinfo {author} {\bibfnamefont {A.}~\bibnamefont {Lipatov}}, \bibinfo {author} {\bibfnamefont {S.}~\bibnamefont {Ryu}}, \bibinfo {author} {\bibfnamefont {D.}~\bibnamefont {Kim}}, \bibinfo {author} {\bibfnamefont {H.}~\bibnamefont {Lee}}, \bibinfo {author} {\bibfnamefont {M.~Y.}\ \bibnamefont {Zhuravlev}}, \bibinfo {author} {\bibfnamefont {C.-B.}\ \bibnamefont {Eom}}, \bibinfo {author} {\bibfnamefont {E.~Y.}\ \bibnamefont {Tsymbal}}, \bibinfo {author} {\bibfnamefont {A.}~\bibnamefont {Sinitskii}}, \ and\ \bibinfo {author} {\bibfnamefont {A.}~\bibnamefont {Gruverman}},\ }\href@noop {} {\bibfield  {journal} {\bibinfo  {journal} {\href{https://doi.org/10.1038/ncomms6518}{\textcolor{blue}{Nat. commun.}}}\ }\textbf {\bibinfo {volume} {\textbf{5}\normalfont(1)}},\ \bibinfo {pages} {5518} (\bibinfo {year} {2014})}\BibitemShut {NoStop}%
\bibitem [{\citenamefont {Zhuravlev}\ \emph {et~al.}(2005)\citenamefont {Zhuravlev}, \citenamefont {Sabirianov}, \citenamefont {Jaswal},\ and\ \citenamefont {Tsymbal}}]{zhuravlev2005giant}%
  \BibitemOpen
  \bibfield  {author} {\bibinfo {author} {\bibfnamefont {M.~Y.}\ \bibnamefont {Zhuravlev}}, \bibinfo {author} {\bibfnamefont {R.~F.}\ \bibnamefont {Sabirianov}}, \bibinfo {author} {\bibfnamefont {S.}~\bibnamefont {Jaswal}}, \ and\ \bibinfo {author} {\bibfnamefont {E.~Y.}\ \bibnamefont {Tsymbal}},\ }\href@noop {} {\bibfield  {journal} {\bibinfo  {journal} {\href{https://doi.org/10.1103/PhysRevLett.94.246802}{\textcolor{blue}{Phys. Rev. Lett.}}}\ }\textbf {\bibinfo {volume} {\textbf{94}\normalfont(24)}},\ \bibinfo {pages} {246802} (\bibinfo {year} {2005})}\BibitemShut {NoStop}%
\bibitem [{\citenamefont {Zhuravlev}\ \emph {et~al.}(2009)\citenamefont {Zhuravlev}, \citenamefont {Wang}, \citenamefont {Maekawa},\ and\ \citenamefont {Tsymbal}}]{Zhuravlev2009}%
  \BibitemOpen
  \bibfield  {author} {\bibinfo {author} {\bibfnamefont {M.~Y.}\ \bibnamefont {Zhuravlev}}, \bibinfo {author} {\bibfnamefont {Y.}~\bibnamefont {Wang}}, \bibinfo {author} {\bibfnamefont {S.}~\bibnamefont {Maekawa}}, \ and\ \bibinfo {author} {\bibfnamefont {E.~Y.}\ \bibnamefont {Tsymbal}},\ }\href@noop {} {\bibfield  {journal} {\bibinfo  {journal} {\href{https://doi.org/10.1063/1.3195075}{\textcolor{blue}{Appl. Phys. Lett.}}}\ }\textbf {\bibinfo {volume} {\textbf{95}\normalfont(5)}},\ \bibinfo {pages} {052902} (\bibinfo {year} {2009})}\BibitemShut {NoStop}%
\bibitem [{\citenamefont {Chang}\ \emph {et~al.}(2017)\citenamefont {Chang}, \citenamefont {Naeemi}, \citenamefont {Nikonov},\ and\ \citenamefont {Gruverman}}]{Chang2017pra}%
  \BibitemOpen
  \bibfield  {author} {\bibinfo {author} {\bibfnamefont {S.-C.}\ \bibnamefont {Chang}}, \bibinfo {author} {\bibfnamefont {A.}~\bibnamefont {Naeemi}}, \bibinfo {author} {\bibfnamefont {D.~E.}\ \bibnamefont {Nikonov}}, \ and\ \bibinfo {author} {\bibfnamefont {A.}~\bibnamefont {Gruverman}},\ }\href@noop {} {\bibfield  {journal} {\bibinfo  {journal} {\href{http://dx.doi.org/10.1103/PhysRevApplied.7.024005}{\textcolor{blue}{Phys. Rev. Appl.}}}\ }\textbf {\bibinfo {volume} {\textbf{7}\normalfont(2)}},\ \bibinfo {pages} {024005} (\bibinfo {year} {2017})}\BibitemShut {NoStop}%
\bibitem [{\citenamefont {Zheng}\ \emph {et~al.}(2022)\citenamefont {Zheng}, \citenamefont {Yang}, \citenamefont {Zhang}, \citenamefont {Li},\ and\ \citenamefont {Liu}}]{zheng2022band}%
  \BibitemOpen
  \bibfield  {author} {\bibinfo {author} {\bibfnamefont {X.}~\bibnamefont {Zheng}}, \bibinfo {author} {\bibfnamefont {Y.}~\bibnamefont {Yang}}, \bibinfo {author} {\bibfnamefont {Q.}~\bibnamefont {Zhang}}, \bibinfo {author} {\bibfnamefont {J.}~\bibnamefont {Li}}, \ and\ \bibinfo {author} {\bibfnamefont {X.}~\bibnamefont {Liu}},\ }\href@noop {} {\bibfield  {journal} {\bibinfo  {journal} {\href{https://doi.org/10.1063/5.0106693}{\textcolor{blue}{ Appl. Phys. Lett.}}}\ }\textbf {\bibinfo {volume} {\textbf{121}\normalfont(13)}},\ \bibinfo {pages} {132902} (\bibinfo {year} {2022})}\BibitemShut {NoStop}%
\bibitem [{\citenamefont {Chang}\ and\ \citenamefont {Xie}(2023{\natexlab{a}})}]{chang2023edl}%
  \BibitemOpen
  \bibfield  {author} {\bibinfo {author} {\bibfnamefont {P.}~\bibnamefont {Chang}}\ and\ \bibinfo {author} {\bibfnamefont {Y.}~\bibnamefont {Xie}},\ }\href@noop {} {\bibfield  {journal} {\bibinfo  {journal} {\href{https://doi.org/10.1109/LED.2022.3225298}{\textcolor{blue}{IEEE Electron Device Lett.}}}\ }\textbf {\bibinfo {volume} {\textbf{44}\normalfont(1)}},\ \bibinfo {pages} {168--171} (\bibinfo {year} {2023}{\natexlab{a}})}\BibitemShut {NoStop}%
\bibitem [{\citenamefont {Hwang}, \citenamefont {Goh},\ and\ \citenamefont {Jeon}(2020)}]{hwang2020effect}%
  \BibitemOpen
  \bibfield  {author} {\bibinfo {author} {\bibfnamefont {J.}~\bibnamefont {Hwang}}, \bibinfo {author} {\bibfnamefont {Y.}~\bibnamefont {Goh}}, \ and\ \bibinfo {author} {\bibfnamefont {S.}~\bibnamefont {Jeon}},\ }\href@noop {} {\bibfield  {journal} {\bibinfo  {journal} {\href{https://doi.org/10.1109/ted.2020.3043728}{\textcolor{blue}{IEEE Trans. on Electron Devices}}}\ }\textbf {\bibinfo {volume} {\textbf{68}\normalfont(2)}},\ \bibinfo {pages} {841--845} (\bibinfo {year} {2020})}\BibitemShut {NoStop}%
\bibitem [{\citenamefont {Joh}, \citenamefont {Jung},\ and\ \citenamefont {Jeon}(2021)}]{joh2021stress}%
  \BibitemOpen
  \bibfield  {author} {\bibinfo {author} {\bibfnamefont {H.}~\bibnamefont {Joh}}, \bibinfo {author} {\bibfnamefont {T.}~\bibnamefont {Jung}}, \ and\ \bibinfo {author} {\bibfnamefont {S.}~\bibnamefont {Jeon}},\ }\href@noop {} {\bibfield  {journal} {\bibinfo  {journal} {\href{https://doi.org/10.1109/TED.2021.3068246}{\textcolor{blue}{IEEE Trans. on Electron Devices}}}\ }\textbf {\bibinfo {volume} {\textbf{68}\normalfont(5)}},\ \bibinfo {pages} {2538--2542} (\bibinfo {year} {2021})}\BibitemShut {NoStop}%
\bibitem [{\citenamefont {Kittel}\ and\ \citenamefont {McEuen}(2018)}]{kittel2018introduction}%
  \BibitemOpen
  \bibfield  {author} {\bibinfo {author} {\bibfnamefont {C.}~\bibnamefont {Kittel}}\ and\ \bibinfo {author} {\bibfnamefont {P.}~\bibnamefont {McEuen}},\ }\href@noop {} {\emph {\bibinfo {title} {Introduction to solid state physics}}}\ (\bibinfo  {publisher} {John Wiley \& Sons},\ \bibinfo {year} {2018})\BibitemShut {NoStop}%
\bibitem [{\citenamefont {Sandu}\ \emph {et~al.}(2022)\citenamefont {Sandu}, \citenamefont {Tibeica}, \citenamefont {Plugaru}, \citenamefont {Nedelcu},\ and\ \citenamefont {Plugaru}}]{sandu2022insights}%
  \BibitemOpen
  \bibfield  {author} {\bibinfo {author} {\bibfnamefont {T.}~\bibnamefont {Sandu}}, \bibinfo {author} {\bibfnamefont {C.}~\bibnamefont {Tibeica}}, \bibinfo {author} {\bibfnamefont {R.}~\bibnamefont {Plugaru}}, \bibinfo {author} {\bibfnamefont {O.}~\bibnamefont {Nedelcu}}, \ and\ \bibinfo {author} {\bibfnamefont {N.}~\bibnamefont {Plugaru}},\ }\href@noop {} {\bibfield  {journal} {\bibinfo  {journal} {\href{https://doi.org/10.3390/nano12101682}{\textcolor{blue}{Nanomaterials}}}\ }\textbf {\bibinfo {volume} {\textbf{12}\normalfont(10)}},\ \bibinfo {pages} {1682} (\bibinfo {year} {2022})}\BibitemShut {NoStop}%
\bibitem [{\citenamefont {Pantel}\ and\ \citenamefont {Alexe}(2010)}]{pantel2010electroresistance}%
  \BibitemOpen
  \bibfield  {author} {\bibinfo {author} {\bibfnamefont {D.}~\bibnamefont {Pantel}}\ and\ \bibinfo {author} {\bibfnamefont {M.}~\bibnamefont {Alexe}},\ }\href@noop {} {\bibfield  {journal} {\bibinfo  {journal} {\href{https://doi.org/10.1103/physrevb.82.134105}{\textcolor{blue}{ Phys. Rev. B}}}\ }\textbf {\bibinfo {volume} {\textbf{82}\normalfont(13)}},\ \bibinfo {pages} {134105} (\bibinfo {year} {2010})}\BibitemShut {NoStop}%
\bibitem [{\citenamefont {Lyu}\ \emph {et~al.}(2019)\citenamefont {Lyu}, \citenamefont {Fina}, \citenamefont {Solanas}, \citenamefont {Fontcuberta},\ and\ \citenamefont {Sánchez}}]{lyu_thickness}%
  \BibitemOpen
  \bibfield  {author} {\bibinfo {author} {\bibfnamefont {J.}~\bibnamefont {Lyu}}, \bibinfo {author} {\bibfnamefont {I.}~\bibnamefont {Fina}}, \bibinfo {author} {\bibfnamefont {R.}~\bibnamefont {Solanas}}, \bibinfo {author} {\bibfnamefont {J.}~\bibnamefont {Fontcuberta}}, \ and\ \bibinfo {author} {\bibfnamefont {F.}~\bibnamefont {Sánchez}},\ }\href@noop {} {\bibfield  {journal} {\bibinfo  {journal} {\href{https://doi.org/10.1021/acsaelm.8b00065}{\textcolor{blue}{ACS Appl. Electron. Mater.}}}\ }\textbf {\bibinfo {volume} {\textbf{1}\normalfont(2)}},\ \bibinfo {pages} {220--228} (\bibinfo {year} {2019})}\BibitemShut {NoStop}%
\bibitem [{\citenamefont {Janovec}(1958)}]{janovec1958theory_ratio}%
  \BibitemOpen
  \bibfield  {author} {\bibinfo {author} {\bibfnamefont {V.}~\bibnamefont {Janovec}},\ }\href@noop {} {\bibfield  {journal} {\bibinfo  {journal} {\href{https://doi.org/10.1007/BF01688741}{\textcolor{blue}{Czechoslovak J. Phys.}}}\ }\textbf {\bibinfo {volume} {8}},\ \bibinfo {pages} {3--15} (\bibinfo {year} {1958})}\BibitemShut {NoStop}%
\bibitem [{\citenamefont {Miller}\ \emph {et~al.}(1990)\citenamefont {Miller}, \citenamefont {Nasby}, \citenamefont {Schwank}, \citenamefont {Rodgers},\ and\ \citenamefont {Dressendorfer}}]{miller1990device}%
  \BibitemOpen
  \bibfield  {author} {\bibinfo {author} {\bibfnamefont {S.}~\bibnamefont {Miller}}, \bibinfo {author} {\bibfnamefont {R.}~\bibnamefont {Nasby}}, \bibinfo {author} {\bibfnamefont {J.}~\bibnamefont {Schwank}}, \bibinfo {author} {\bibfnamefont {M.}~\bibnamefont {Rodgers}}, \ and\ \bibinfo {author} {\bibfnamefont {P.}~\bibnamefont {Dressendorfer}},\ }\href@noop {} {\bibfield  {journal} {\bibinfo  {journal} {\href{https://doi.org/10.1063/1.346845}{\textcolor{blue}{J. Appl. Phys.}}}\ }\textbf {\bibinfo {volume} {\textbf{12}\normalfont(23)}},\ \bibinfo {pages} {6463--6471} (\bibinfo {year} {1990})}\BibitemShut {NoStop}%
\bibitem [{\citenamefont {Sutor}, \citenamefont {Rupitsch},\ and\ \citenamefont {Lerch}(2010)}]{sutor2010preisach}%
  \BibitemOpen
  \bibfield  {author} {\bibinfo {author} {\bibfnamefont {A.}~\bibnamefont {Sutor}}, \bibinfo {author} {\bibfnamefont {S.~J.}\ \bibnamefont {Rupitsch}}, \ and\ \bibinfo {author} {\bibfnamefont {R.}~\bibnamefont {Lerch}},\ }\href@noop {} {\bibfield  {journal} {\bibinfo  {journal} {\href{https://doi.org/10.1007/s00339-010-5884-9}{\textcolor{blue}{Appl. Phys. A}}}\ }\textbf {\bibinfo {volume} {\textbf{100}\normalfont(2)}},\ \bibinfo {pages} {425--430} (\bibinfo {year} {2010})}\BibitemShut {NoStop}%
\bibitem [{\citenamefont {Mayergoyz}\ and\ \citenamefont {Friedman}(1988)}]{mayergoyz1988generalized}%
  \BibitemOpen
  \bibfield  {author} {\bibinfo {author} {\bibfnamefont {I.~D.}\ \bibnamefont {Mayergoyz}}\ and\ \bibinfo {author} {\bibfnamefont {G.}~\bibnamefont {Friedman}},\ }\href@noop {} {\bibfield  {journal} {\bibinfo  {journal} {\href{https://doi.org/10.1109/20.43892}{\textcolor{blue}{IEEE trans. on Magnetics}}}\ }\textbf {\bibinfo {volume} {\textbf{24}\normalfont(1)}},\ \bibinfo {pages} {212--217} (\bibinfo {year} {1988})}\BibitemShut {NoStop}%
\bibitem [{\citenamefont {Ando}\ and\ \citenamefont {Itoh}(1987)}]{ando1987calculation}%
  \BibitemOpen
  \bibfield  {author} {\bibinfo {author} {\bibfnamefont {Y.}~\bibnamefont {Ando}}\ and\ \bibinfo {author} {\bibfnamefont {T.}~\bibnamefont {Itoh}},\ }\href@noop {} {\bibfield  {journal} {\bibinfo  {journal} {\href{https://doi.org/10.1063/1.338082}{\textcolor{blue}{J. Appl. Phys.}}}\ }\textbf {\bibinfo {volume} {\textbf{61}\normalfont(4)}},\ \bibinfo {pages} {1497--1502} (\bibinfo {year} {1987})}\BibitemShut {NoStop}%
\bibitem [{\citenamefont {Zhou}\ \emph {et~al.}(2021)\citenamefont {Zhou}, \citenamefont {Jiao}, \citenamefont {Zhou}, \citenamefont {Kong}, \citenamefont {Luo}, \citenamefont {Sun}, \citenamefont {Zheng}, \citenamefont {Wang}, \citenamefont {Zhang}, \citenamefont {Liu} \emph {et~al.}}]{zhou2021time}%
  \BibitemOpen
  \bibfield  {author} {\bibinfo {author} {\bibfnamefont {Z.}~\bibnamefont {Zhou}}, \bibinfo {author} {\bibfnamefont {L.}~\bibnamefont {Jiao}}, \bibinfo {author} {\bibfnamefont {J.}~\bibnamefont {Zhou}}, \bibinfo {author} {\bibfnamefont {Q.}~\bibnamefont {Kong}}, \bibinfo {author} {\bibfnamefont {S.}~\bibnamefont {Luo}}, \bibinfo {author} {\bibfnamefont {C.}~\bibnamefont {Sun}}, \bibinfo {author} {\bibfnamefont {Z.}~\bibnamefont {Zheng}}, \bibinfo {author} {\bibfnamefont {X.}~\bibnamefont {Wang}}, \bibinfo {author} {\bibfnamefont {D.}~\bibnamefont {Zhang}}, \bibinfo {author} {\bibfnamefont {G.}~\bibnamefont {Liu}},  \emph {et~al.},\ }\href@noop {} {\bibfield  {journal} {\bibinfo  {journal} {\href{https://doi.org/10.1109/led.2021.3128998}{\textcolor{blue}{IEEE Electron Device Lett.}}}\ }\textbf {\bibinfo {volume} {\textbf{43}\normalfont(1)}},\ \bibinfo {pages} {158--161} (\bibinfo {year} {2021})}\BibitemShut {NoStop}%
\bibitem [{\citenamefont {Datta}(2005)}]{datta2005quantum}%
  \BibitemOpen
  \bibfield  {author} {\bibinfo {author} {\bibfnamefont {S.}~\bibnamefont {Datta}},\ }\href@noop {} {}\ (\bibinfo  {publisher} {Cambridge university press},\ \bibinfo {year} {2005})\BibitemShut {NoStop}%
\bibitem [{\citenamefont {Sharma}, \citenamefont {Tulapurkar},\ and\ \citenamefont {Muralidharan}(2016)}]{sharma2016ultrasensitive}%
  \BibitemOpen
  \bibfield  {author} {\bibinfo {author} {\bibfnamefont {A.}~\bibnamefont {Sharma}}, \bibinfo {author} {\bibfnamefont {A.}~\bibnamefont {Tulapurkar}}, \ and\ \bibinfo {author} {\bibfnamefont {B.}~\bibnamefont {Muralidharan}},\ }\href@noop {} {\bibfield  {journal} {\bibinfo  {journal} {\href{https://doi.org/10.1109/TED.2016.2606354}{\textcolor{blue}{IEEE Trans. on Electron Devices}}}\ }\textbf {\bibinfo {volume} {\textbf{63}\normalfont(11)}},\ \bibinfo {pages} {4527--4534} (\bibinfo {year} {2016})}\BibitemShut {NoStop}%
\bibitem [{\citenamefont {Fan}\ \emph {et~al.}(2016)\citenamefont {Fan}, \citenamefont {Xiao}, \citenamefont {Wang}, \citenamefont {Zhang}, \citenamefont {Deng}, \citenamefont {Liu}, \citenamefont {Dong}, \citenamefont {Wang},\ and\ \citenamefont {Chen}}]{fan2016ferroelectricity_hzo}%
  \BibitemOpen
  \bibfield  {author} {\bibinfo {author} {\bibfnamefont {Z.}~\bibnamefont {Fan}}, \bibinfo {author} {\bibfnamefont {J.}~\bibnamefont {Xiao}}, \bibinfo {author} {\bibfnamefont {J.}~\bibnamefont {Wang}}, \bibinfo {author} {\bibfnamefont {L.}~\bibnamefont {Zhang}}, \bibinfo {author} {\bibfnamefont {J.}~\bibnamefont {Deng}}, \bibinfo {author} {\bibfnamefont {Z.}~\bibnamefont {Liu}}, \bibinfo {author} {\bibfnamefont {Z.}~\bibnamefont {Dong}}, \bibinfo {author} {\bibfnamefont {J.}~\bibnamefont {Wang}}, \ and\ \bibinfo {author} {\bibfnamefont {J.}~\bibnamefont {Chen}},\ }\href@noop {} {\bibfield  {journal} {\bibinfo  {journal} {\href{https://doi.org/10.1063/1.4953461} {\textcolor{blue}{Appl. Phys. Lett.}}}\ }\textbf {\bibinfo {volume} {\textbf{108}\normalfont(23)}},\ \bibinfo {pages} {232905} (\bibinfo {year} {2016})}\BibitemShut {NoStop}%
\bibitem [{\citenamefont {Robertson}(2013)}]{robertson2013band}%
  \BibitemOpen
  \bibfield  {author} {\bibinfo {author} {\bibfnamefont {J.}~\bibnamefont {Robertson}},\ }\href@noop {} {\bibfield  {journal} {\bibinfo  {journal} {\href{https://doi.org/10.1116/1.4818426}{\textcolor{blue}{J. Vac. Sci. Technol. A}}}\ }\textbf {\bibinfo {volume} {\textbf{31}\normalfont(5)}},\ \bibinfo {pages} {050821} (\bibinfo {year} {2013})}\BibitemShut {NoStop}%
\bibitem [{\citenamefont {Fillot}\ \emph {et~al.}(2005)\citenamefont {Fillot}, \citenamefont {Morel}, \citenamefont {Minoret}, \citenamefont {Matko}, \citenamefont {Ma{\^\i}trejean}, \citenamefont {Guillaumot}, \citenamefont {Chenevier},\ and\ \citenamefont {Billon}}]{fillot2005investigations}%
  \BibitemOpen
  \bibfield  {author} {\bibinfo {author} {\bibfnamefont {F.}~\bibnamefont {Fillot}}, \bibinfo {author} {\bibfnamefont {T.}~\bibnamefont {Morel}}, \bibinfo {author} {\bibfnamefont {S.}~\bibnamefont {Minoret}}, \bibinfo {author} {\bibfnamefont {I.}~\bibnamefont {Matko}}, \bibinfo {author} {\bibfnamefont {S.}~\bibnamefont {Ma{\^\i}trejean}}, \bibinfo {author} {\bibfnamefont {B.}~\bibnamefont {Guillaumot}}, \bibinfo {author} {\bibfnamefont {B.}~\bibnamefont {Chenevier}}, \ and\ \bibinfo {author} {\bibfnamefont {T.}~\bibnamefont {Billon}},\ }\href@noop {} {\bibfield  {journal} {\bibinfo  {journal} {\href{https://doi.org/10.1016/j.mee.2005.07.083}{\textcolor{blue}{Microelectron. Eng.}}}\ }\textbf {\bibinfo {volume} {\textbf{82}\normalfont(3-4)}},\ \bibinfo {pages} {248--253} (\bibinfo {year} {2005})}\BibitemShut {NoStop}%
\bibitem [{\citenamefont {Chang}\ and\ \citenamefont {Xie}(2023{\natexlab{b}})}]{chang2023ted}%
  \BibitemOpen
  \bibfield  {author} {\bibinfo {author} {\bibfnamefont {P.}~\bibnamefont {Chang}}\ and\ \bibinfo {author} {\bibfnamefont {Y.}~\bibnamefont {Xie}},\ }\href@noop {} {\bibfield  {journal} {\bibinfo  {journal} {\href{https://doi.org/10.1109/TED.2023.3251958}{\textcolor{blue}{IEEE Trans. on Electron Devices}}}\ }\textbf {\bibinfo {volume} {\textbf{70}\normalfont(5)}},\ \bibinfo {pages} {2282--2290} (\bibinfo {year} {2023}{\natexlab{b}})}\BibitemShut {NoStop}%
\bibitem [{\citenamefont {Materlik}, \citenamefont {K{\"u}nneth},\ and\ \citenamefont {Kersch}(2015)}]{materlik2015origin}%
  \BibitemOpen
  \bibfield  {author} {\bibinfo {author} {\bibfnamefont {R.}~\bibnamefont {Materlik}}, \bibinfo {author} {\bibfnamefont {C.}~\bibnamefont {K{\"u}nneth}}, \ and\ \bibinfo {author} {\bibfnamefont {A.}~\bibnamefont {Kersch}},\ }\href@noop {} {\bibfield  {journal} {\bibinfo  {journal} {\href{http://dx.doi.org/10.1063/1.4916707}{\textcolor{blue}{J. Appl. Phys.}}}\ }\textbf {\bibinfo {volume} {\textbf{117}\normalfont(13)}},\ \bibinfo {pages} {134109} (\bibinfo {year} {2015})}\BibitemShut {NoStop}%
\bibitem [{\citenamefont {Mondal}, \citenamefont {Kumar}\ \emph {et~al.}(2016)\citenamefont {Mondal}, \citenamefont {Kumar} \emph {et~al.}}]{mondal2016tunable}%
  \BibitemOpen
  \bibfield  {author} {\bibinfo {author} {\bibfnamefont {S.}~\bibnamefont {Mondal}}, \bibinfo {author} {\bibfnamefont {A.}~\bibnamefont {Kumar}},  \emph {et~al.},\ }\href@noop {} {\bibfield  {journal} {\bibinfo  {journal} {\href{https://doi.org/10.1016/j.spmi.2016.10.054}{\textcolor{blue}{Superlattices and Microstruct.}}}\ }\textbf {\bibinfo {volume} {100}},\ \bibinfo {pages} {876--885} (\bibinfo {year} {2016})}\BibitemShut {NoStop}%
\bibitem [{\citenamefont {Tripura~Sundari}\ \emph {et~al.}(2014)\citenamefont {Tripura~Sundari}, \citenamefont {Ramaseshan}, \citenamefont {Jose}, \citenamefont {Dash},\ and\ \citenamefont {Tyagi}}]{tripura2014investigation}%
  \BibitemOpen
  \bibfield  {author} {\bibinfo {author} {\bibfnamefont {S.}~\bibnamefont {Tripura~Sundari}}, \bibinfo {author} {\bibfnamefont {R.}~\bibnamefont {Ramaseshan}}, \bibinfo {author} {\bibfnamefont {F.}~\bibnamefont {Jose}}, \bibinfo {author} {\bibfnamefont {S.}~\bibnamefont {Dash}}, \ and\ \bibinfo {author} {\bibfnamefont {A.}~\bibnamefont {Tyagi}},\ }\href@noop {} {\bibfield  {journal} {\bibinfo  {journal} {\href{https://doi.org/10.1063/1.4862485}{\textcolor{blue}{J. Appl. Phys.}}}\ }\textbf {\bibinfo {volume} {\textbf{115}\normalfont(3)}} (\bibinfo {year} {2014})}\BibitemShut {NoStop}%
\bibitem [{\citenamefont {Monaghan}\ \emph {et~al.}(2009)\citenamefont {Monaghan}, \citenamefont {Hurley}, \citenamefont {Cherkaoui}, \citenamefont {Negara},\ and\ \citenamefont {Schenk}}]{monaghan2009determination}%
  \BibitemOpen
  \bibfield  {author} {\bibinfo {author} {\bibfnamefont {S.}~\bibnamefont {Monaghan}}, \bibinfo {author} {\bibfnamefont {P.}~\bibnamefont {Hurley}}, \bibinfo {author} {\bibfnamefont {K.}~\bibnamefont {Cherkaoui}}, \bibinfo {author} {\bibfnamefont {M.}~\bibnamefont {Negara}}, \ and\ \bibinfo {author} {\bibfnamefont {A.}~\bibnamefont {Schenk}},\ }\href@noop {} {\bibfield  {journal} {\bibinfo  {journal} {\href{https://doi.org/10.1016/j.sse.2008.09.018}{\textcolor{blue}{Solid-State Electronics}}}\ }\textbf {\bibinfo {volume} {\textbf{53}\normalfont(4)}},\ \bibinfo {pages} {438--444} (\bibinfo {year} {2009})}\BibitemShut {NoStop}%
\bibitem [{\citenamefont {Gulomov}\ \emph {et~al.}(2023)\citenamefont {Gulomov}, \citenamefont {Accouche}, \citenamefont {Aliev}, \citenamefont {Ghandour},\ and\ \citenamefont {Gulomova}}]{gulomov2023investigation}%
  \BibitemOpen
  \bibfield  {author} {\bibinfo {author} {\bibfnamefont {J.}~\bibnamefont {Gulomov}}, \bibinfo {author} {\bibfnamefont {O.}~\bibnamefont {Accouche}}, \bibinfo {author} {\bibfnamefont {R.}~\bibnamefont {Aliev}}, \bibinfo {author} {\bibfnamefont {R.}~\bibnamefont {Ghandour}}, \ and\ \bibinfo {author} {\bibfnamefont {I.}~\bibnamefont {Gulomova}},\ }\href@noop {} {\bibfield  {journal} {\bibinfo  {journal} {{\href{http://dx.doi.org/10.1109/ACCESS.2023.3268033}{\textcolor{blue}IEEE Access}}}\ }\textbf {\bibinfo {volume} {11}},\ \bibinfo {pages} {38970--38981} (\bibinfo {year} {2023})}\BibitemShut {NoStop}%
\bibitem [{\citenamefont {Abuwasib}\ \emph {et~al.}(2016)\citenamefont {Abuwasib}, \citenamefont {Lu}, \citenamefont {Li}, \citenamefont {Buragohain}, \citenamefont {Lee}, \citenamefont {Eom}, \citenamefont {Gruverman},\ and\ \citenamefont {Singisetti}}]{abuwasib2016scaling}%
  \BibitemOpen
  \bibfield  {author} {\bibinfo {author} {\bibfnamefont {M.}~\bibnamefont {Abuwasib}}, \bibinfo {author} {\bibfnamefont {H.}~\bibnamefont {Lu}}, \bibinfo {author} {\bibfnamefont {T.}~\bibnamefont {Li}}, \bibinfo {author} {\bibfnamefont {P.}~\bibnamefont {Buragohain}}, \bibinfo {author} {\bibfnamefont {H.}~\bibnamefont {Lee}}, \bibinfo {author} {\bibfnamefont {C.-B.}\ \bibnamefont {Eom}}, \bibinfo {author} {\bibfnamefont {A.}~\bibnamefont {Gruverman}}, \ and\ \bibinfo {author} {\bibfnamefont {U.}~\bibnamefont {Singisetti}},\ }\href@noop {} {\bibfield  {journal} {\bibinfo  {journal} {\href{https://doi.org/10.1063/1.4947020}{\textcolor{blue}{Appl. Phys. Lett.}}}\ }\textbf {\bibinfo {volume} {\textbf{108}\normalfont(15)}},\ \bibinfo {pages} {152904} (\bibinfo {year} {2016})}\BibitemShut {NoStop}%
\end{thebibliography}%
\end{document}